# Determination of I-V Curves of HTS Tapes From The Frequency Dependent AC Transport Loss

Eduard Demenčík, Aurelien Godfrin, Víctor Zermeño and Francesco Grilli

*Abstract*— The current-voltage characteristics of superconductors are commonly determined by means of a standard 4-probe DC measurement technique. In coated conductors, however, one often encounters problems due to weak or no stabilization and in such case, when the risk of burning the tape is high, the 4-probe approach can fail. In other cases such as in HTS cable structures, besides vulnerability and high risk of damage, it may also be difficult to achieve a high critical current level with a power supply that is normally used to characterize single strands. We therefore investigated an alternative method to determine the DC *I-V* curve, based on measuring the AC transport loss as a function of frequency in overcritical current range. In the AC regime, both the hysteretic loss component and the joule dissipation are present for applied transport currents exceeding the critical current $I_c$. In this work, AC loss measurements at different frequencies are related to different electric fields in DC and a curve similar to the DC *I-V* is obtained. From this curve, the power index $N$ and $I_c$ values are estimated. AC regime operation can be more beneficial for two reasons: firstly, because very high currents can be achieved easily by using a transformer; secondly, because the effective joule heat dissipation with overcritical AC currents is much lower than that with DC of the same amplitudes. In the experiment, we used a 4 mm wide YBCO reference tape with 20 μm of Cu-stabilization. For the comparison with AC transport loss, the sample was measured with the standard 4-probe technique and remained stable even in overload condition. To support our study, numerical simulations, reproducing both transient and stationary operation conditions, were performed. A similar relation between the AC current transport loss and the *I-V* curves was also observed in the simulations.

## I. Introduction

HIGH TC COATED CONDUCTORS (CCs) for AC applications can be modelled based on the static characterization in DC regime. In DC regime, the heat dissipated in the superconductor fed with current below $I_c$ is negligible. When the amplitude of the current exceeds the critical value then the losses increase very quickly and the thermal heating becomes considerable. This heating furthermore tends to reduce the current carrying capacity. The reason for this is the very steep power low dependence of the electric field on the applied current $E \sim J^N$. Thermal and electrical stability of the CC tapes is therefore an issue for further use in AC applications such as cables and magnets, where excessive heating can be present while AC operation or ramping of the magnets. It has been shown in previous works [1], [2] that, whereas for the well stabilized BSCCO tapes, the transport current can overcome the $I_c$ value by several times repeatedly without damage, keeping the safety margin in the case of the CCs is crucial. In order to characterize the superconductor, the standard DC *I-V* measurements are done. In addition to the standard DC characterization, the use of AC for the determination of the current carrying capability of superconductors seems an option with some benefits, as well. It may be used in the case when a high risk of damage in HTS (tape, cable, etc.) structure is possible. This is due to the fact that the heat dissipation in the AC case is lower when the same amplitude of current as in the DC case is applied. Another advantage of the use of AC is the flexibility in scaling of the AC current amplitude with help of a transformer. Especially it can be useful in HTS cables, where the DC current amplitudes are rather high and the required AC level can be reached even with using a relatively small AC power supply. However, the use of AC seems inadequate because of the involved induced flux creep voltage. In the work of Gomory et al. [1] it was shown that the total transport AC loss can be obtained as a sum of magnetic and resistive loss component in the whole range of AC currents. In addition to that, it was experimentally proved that the magnetic loss component saturates at $I_a = I_c$. This allows for a new AC loss data interpretation in the overload range ($I_a > I_c$) with the help of the appropriate effective current calculation. In literature, the saturation of the current carrying capacity (defined by the current amplitude at which the above mentioned magnetic loss is saturated) of the CC tape as a function of frequency was observed. This current is referred to as frequency dependent critical current $I_c$, as proposed by Thakur et al. [3]. They showed the dependence of the $I_c$ increasing as $f^{1/n}$ with frequency. Although the mechanism of the measured frequency dependence of $I_c$ on frequency is not very clear, one possible interpretation of this dependence is that it might be a consequence of the thermally activated creep of vortices. Frequency dependence of $I_c$ was a subject of quite extensive studies, and the references can be found within [3]. In this article we test the hypothesis that the observed frequency dependence of the "$I_c$" is a consequence of the *E-J* relationship. In order to do so, we compared the AC transport loss curves obtained for various frequencies with the power line obtained from the DC *I-V* measurements. The

Manuscript received September 8, 2015.
This work was partially supported by the Helmholtz-University Young Investigator Group Grant VH-NG-617.
E. Demenčík, A. Godfrin, V. Zermeño and F. Grilli are with Institute for Technical Physics, Karlsruhe Institute of Technology, 76344 Germany.
Corresponding author's: phone: +49721608-28067; fax: +49721608-22849; e-mail: eduard.demencik@kit.edu.

experimental data are supported by analytical Norris formula for strip scaled with frequency and critical current.

## II. METHODOLOGY

For the experiments a 4 mm YBCO sample with 20 µm of copper stabilization manufactured by SuperPower, Inc., was used. The CC sample was characterized by a standard 4-probe method and the critical current was determined for the electric field criterion of $E_c = 10^{-4}$ V.m$^{-1}$. The measured CC sample was supplied with the harmonic AC current $I(t) = I_a \cdot \sin(2 \cdot \pi \cdot f \cdot t)$ with frequencies covering nearly two orders of magnitude from 2 Hz up to 144 Hz. In this range, the transport AC loss was measured as a function of frequency. The resistive part of the measured AC loss curves was compared with the DC current-voltage curve obtained by measurement in the self-field condition. To do this correctly, the current in the resistive part of the AC loss curve was recalculated with the newly introduced effective values. The AC loss experimental data was supported by the analytical Norris formula for a thin strip.

## III. RESULTS AND DISCUSSION

### A. DC characterization in self-field

The 60 mm long piece of sample was characterized and the DC current-voltage characteristic was obtained by using a 4-probe technique, see In Fig. 1a). In Fig. 1b), the DC-power line is plotted.

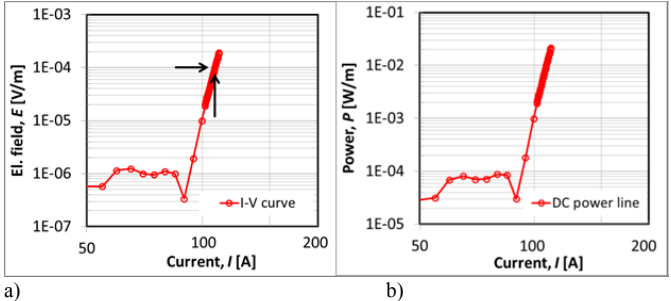

a)　　　　　　　　　　　　　b)
Figure 1. a) Current – voltage characteristics of a 4 mm wide copper stabilized CC tape at zero external field, $I_c$ = 108A
b) Power line obtained as $P = E \cdot I = f(I)$

This line represents the power dissipation of a superconductor in the DC regime derived from the *I-V* curve measurement.
Data hence represent the power law resistive dissipation (note the noise to the left from the steady slope comes from the *I-V* measurement):

$$P = E \cdot I, \text{ where } E = E_c \cdot (I / I_c)^N \quad (1)$$

### B. AC transport loss

A CC 4 mm wide and 0.3 m long piece of sample was supplied with a harmonic alternating current with amplitudes exceeding the critical current value. The AC transport loss was measured with the integration technique and the power dissipation per unit length, as a function of frequency of the AC current, is shown in Fig. 2.

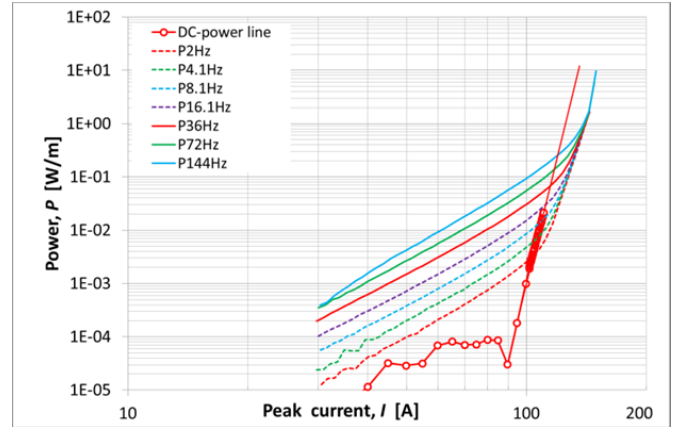

Figure 2. AC transport loss measured at frequency range from 2 Hz up to 144 Hz in comparison with the DC – power line from fig. 1b).

From the measurements, as well as from the previous works published [1]–[5], the curves can be split into two regions. The low current region on the left corresponds to the superconductor's hysteresis (magnetic) loss region, whereas the high current region on the right defines the resistive behavior in overload, which corresponds to the power-law, non-linear resistance of the CC tape. For direct comparison, the DC power line of Fig. 1b is plotted in red circles in Fig. 2. The hysteretic loss in the current range $I_a < I_c$ is associated with the movement of vortices inside the CC tape. At the current applied equal to the critical current amplitude, the flux creep and flux flow region occur. Superconducting hysteretic behavior saturates and the resistive loss described by the power law $E \sim J^N$ starts to dominate. In this region ($I_a > I_c$) the DC dissipation should match the AC measured curves. The discrepancy seen in Fig. 3 is discussed later in the paragraph on the effective overload current calculation.

### C. Evaluation of the effective overload current

In Fig. 2, the discrepancy between the DC "power line" and the resistive regime in the AC transport loss is observable. In the present work we explain this by introducing a calculation of an effective value of the current in overload amplitudes. This model is based on the assumption that the total loss is an algebraic sum of both the magnetic and the resistive AC loss components [1]. Furthermore, the magnetic component saturates when $I_a = I_c$. Such behavior was suggested and experimentally proved by Gomory et al. [1]. Hence in the amplitudes of the applied current below the critical current ($I_a < I_c$) the magnetic loss component is dominant compared to the resistive one which is practically zero. In the overload current range above the $I_c$, the resistive part takes over very quickly as the magnetic component saturates at the $I_a = I_c$ amplitude, see Fig. 4 within the Ref. [1].

It is well known that in normal metals the transport AC current creates joule dissipation and an effective value of the AC current (RMS) exists. By definition, at this current level, the same amount of heat energy as in the DC condition is dissipated, see [7]. However, there is no such an effective value for the case of superconductors due to the different dissipative process in AC and in DC regime. In AC regime, the loss is related to the movement of vortices which yields measurable AC loss contribution. Whereas the loss in the DC is virtually zero in subcritical current range $I_a < I_c$, so no fair comparison between DC and AC can be made in this range of amplitudes ($I_a < I_c$). However, when the CC tape is exposed to the currents exceeding the tape's capacity defined by the critical current, the superconductor dissipates energy not only in the hysteretic way, but also by joule dissipation, same as a non-linear resistor with the power-law dependence, Eq. (1).

In Fig. 3, the range of amplitudes, in which hysteretic and resistive regime is dominant, is shown, similar to earlier reported in [5], Eq. (12). Since the hysteresis loss component is saturated when $I_a = I_c$ this means that the overload current generates the joule dissipation on top of that, only. Which leads to the fact that the AC loss measured at amplitudes exceeding the $I_c$ value should be no longer related to the amplitude of the applied AC current, but rather to an effective value of the applied current. This value can be calculated as a sum of the saturation current value and the RMS value ($I = I_c + I_{eff}$) of the overload current (marked in green line in Fig. 3) over the whole period.

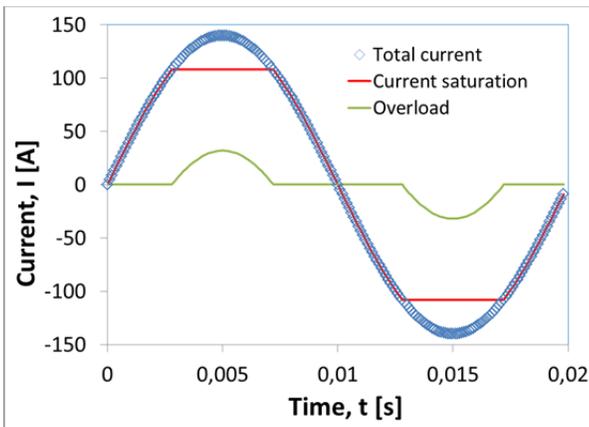

Figure 3. Calculation of the effective value in overload currents range.

The effective value of the plotted green signal $i(t)$ is determined by definition as the integral:

$$I_{eff} = \sqrt{\frac{1}{T} \cdot \int_0^T i^2(t) \cdot dt} \qquad (2)$$

After calculation of the effective values of the current data points, represented on *x*-axis on Fig. 4, the data show nice agreement in the resistive part with the measured DC power line, marked in red circles.

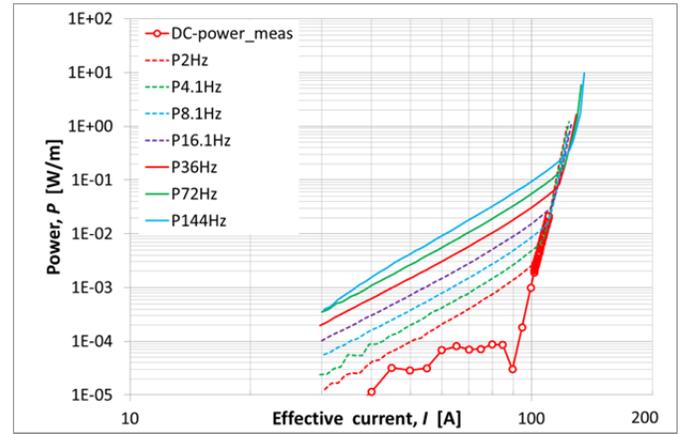

Figure 4. AC loss with calculated effective current values in overload amplitudes (thin dotted lines represent Norris strip with varying $I_c$ parameter).

### D. Model calculations and saturation power line

Based on the discussion carried out above, it is possible to distinguish two regions in which different AC loss mechanisms prevail. The subtle division line between them is defined by the saturation of the current in superconductor. This is the case when the current amplitude, higher than a certain threshold value, is applied. On the power plot, this line can be obtained using two asymptotic lines, as shown in Fig. 5 (and as suggested, for FEM simulations, by Stavrev et al. in [8]). An alternative way is by calculating the power loss with the help of Norris's formula and scaling the loss in one cycle by the frequency, as follows:

$$P/l = Q_s/l \cdot f \qquad (3)$$

where the loss per cycle of a thin strip is given by [9]:

$$Q_s = \mu_0 \cdot I_c^2 /\pi \cdot [(1+F)\cdot\ln(1+F) - F^2 + (1-F)\cdot\ln(1-F)] \qquad (4)$$
$$F = I / I_c$$

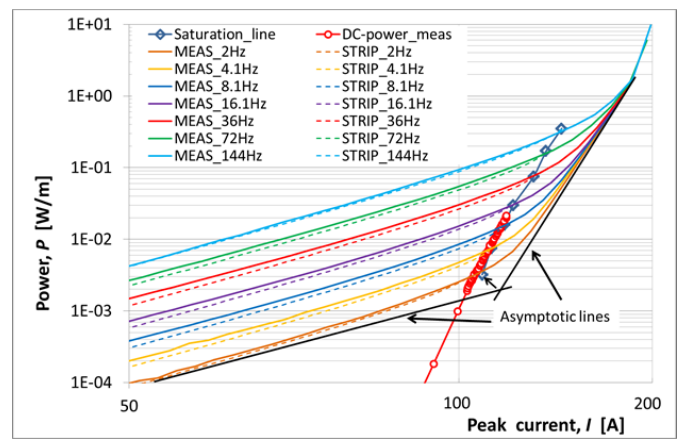

Figure 5. AC-loss – definition of the "saturation line" with the help of assymptotic lines from lower and higher range for each frequency curve (black lines correspond to f = 2 Hz). The dashed lines show loss evaluated with Norris strip formula, scaled with frequency. The $I_c$ parameter in the Norris formula was scaled for each frequency so that it matches the diamond symbols in the plot. For comparison, in red circles, the DC power line is shown.

The $I_c$ parameter in the equation is scaled such that the model AC loss curve fits well the measured data. This is done separately for each frequency. The power line is determined as the series of the $I_c$ values found for each frequency. These values define the current saturation threshold level for each frequency. These values are listed in tab. 1 (represented by blue diamonds in Fig. 5).

Tab. 1.

| f [Hz] | 2 | 4 | 8 | 16 | 36 | 72 | 144 |
|---|---|---|---|---|---|---|---|
| $I_{saturation}$ [A] | 103 | 107 | 110 | 112 | 117 | 120 | 124 |

With respect to these values the effective overload currents are calculated, for each frequency individually. Note that the line depicting the saturation of the current at each frequency, marked as blue diamonds in Fig. 5, is almost in perfect agreement with the power line obtained from the DC measurement. And taking the effective current in overload range into account we were finally able to match the resistive part of the AC loss curves with the DC power line, as originally expected. And moreover, since the power dissipation in the resistive part in AC was found equivalent with the one in DC, additionally, the DC current-voltage relationship can be easily extracted from the AC measurements.

## IV. CONCLUSIONS

We investigated AC transport loss of a 4 mm wide CC sample with 20 μm of copper stabilization. The sample was characterized by a standard 4 probe method and DC I-V curve was measured. The critical current was determined by the electric field criterion of $10^{-4}$ V.m$^{-1}$. The AC loss in overload conditions, as a function of frequency was measured by the integration technique. These measured AC loss curves were compared with the curve derived from the DC measured *I-V* characteristics, referred to as the "power line", which represents the heat dissipation of the tape in the DC regime. The AC and the DC data are then directly compared. The discrepancy in the resistive part of the AC dissipation curves, compared to the DC power line, is observed. In contrast to the other approaches, we suggested to leave the power dissipation, measured as a physical quantity, intact and rather recalculate the amplitude of the current. The reasons for this are: first, the separation of the resistive and the magnetic loss component from the measured power is not so straightforward. Second, the approach does not allow for the direct comparison of the DC and the AC measurements. Thus, the advantage of the suggested new procedure is the direct comparison of the dissipation of the CC tape in DC with the AC. In which case, the DC *I-V* curve can be extracted from the AC loss measurement. In order to do so correctly, we introduce the effective current, calculated for the amplitudes of the current in the over-load condition. The effective values of the current replace the amplitude values in the range where magnetic loss is saturated. As we have shown a fair comparison of the AC and the DC power dissipation is possible in the resistive part (overload conditions) of the AC loss curve. As a consequence of these presented results, it is hence possible to reconstruct the *I-V* curve from the AC transport loss measurements.

ACKNOWLEDGMENT

This work was partially supported by the Helmholtz-University Young Investigator Group Grant VH-NG-617.